\documentclass[prl,aps,showpacs,superscriptaddress,twocolumn]{revtex4}
\usepackage{color} 
\usepackage{bm}
\usepackage{xcolor}
\usepackage{dcolumn}
\usepackage{times}
\usepackage{amssymb} 
\usepackage{amsmath} 
\usepackage{ragged2e}
\usepackage{graphicx}
\usepackage{epstopdf}
\usepackage[english]{babel}
\usepackage{mathrsfs}
\usepackage{textcomp}
\usepackage{amsthm}
\usepackage{amsfonts}
\usepackage{soul}
\usepackage{braket} 
\usepackage{hyperref}
\usepackage{cleveref}
\usepackage{float}
\newcommand{\be}{\begin{equation}}
\newcommand{\ee}{\end{equation}}
\newcommand{\bea}{\begin{eqnarray}}
\newcommand{\eea}{\end{eqnarray}}
\newcommand{\ba}{\begin{array}}
\newcommand{\ea}{\end{array}}

\begin{document}

\title{Interaction induced fractionalization and topological superconductivity in the polar molecules anisotropic $t-J$ model}

\author{Serena Fazzini}
\affiliation{Institute for condensed matter physics and 
complex systems, DISAT, Politecnico di Torino, I-10129, Italy}
\affiliation{Department of Physics  and Research Center OPTIMAS, University of Kaiserslautern, D-67663 Kaiserslautern, Germany}
\author{Luca Barbiero}
\affiliation{Center for Nonlinear Phenomena and Complex Systems,
Universit\'e Libre de Bruxelles, CP 231, Campus Plaine, B-1050 Brussels, Belgium}
\author{Arianna Montorsi}
\affiliation{Institute for condensed matter physics and 
complex systems, DISAT, Politecnico di Torino, I-10129, Italy}

\date{November 30, 2018}

\begin{abstract}
We show that the interplay between antiferromagnetic interaction and hole motion gives rise to a topological superconducting phase. This is captured by the one dimensional anisotropic $t-J$ model which can be experimentally achieved with ultracold polar molecules trapped onto an optical lattice. As a function of the anisotropy strength we find that different quantum phases appear, ranging from a gapless Luttinger liquid to spin gapped conducting and superconducting regimes. In presence of appropriate $z$-anisotropy, we also prove that a phase characterized by non-trivial topological order takes place. The latter is described uniquely by a finite non local string parameter and presents robust edge spin fractionalization. These results allow to explore quantum phases of matter where topological superconductivity is induced by the interaction.
\end{abstract}

\maketitle

\paragraph{Introduction.}
Topological quantum matter has recently attracted huge interest from different research fields \cite{Asorey,Goldman_topology,Ozawa_review}. In this context the presence of gapless edge modes associated to a gapped bulk \cite{hasan,qi} can give rise to unique properties like quantized conductance \cite{klitzking,laughlin,thouless,moore} and charge fractionalization \cite{tsui,laughlin1,su}. Thanks to symmetry arguments, a full understanding of the aforementioned features can be obtained for non interacting systems allowing to classify the so called topological insulators \cite{kane,bernevig} and superconductors \cite{AlZi,Schal}. 
Crucially this approach becomes unstable in presence of interaction\cite{Kit} and the concept of symmetry protection can be exploited to still classify topological phases\cite{Wen, PTBO}. In particular it has been proved\cite{MDIR} that, for strongly correlated systems, the appearance of protected localized edge states is identified by a finite value of a nonlocal string order parameter \cite{Nijs1989}. Furthermore the latter captures the hidden antiferromagnetic ordering of some degrees of freedom (for instance, $\uparrow$ and $\downarrow$ states of a spin $1$ model), diluted in the background of the others (for instance, $0$ state). 
Celebrated example of this is the Haldane phase characteristic of several interacting one dimensional models \cite{Haldane1983,aklt,dallatorre,nonne, dalmonte,kobayashi,barbiero1,DoMo,cohen,pohl,fazzini, barbiero2}. Noticeably these studies all focus on hidden antiferromagnetism in presence of a gapped charge channel, thus describing topological insulating regimes. Therefore finding microscopic interacting Hamiltonians supporting the presence of non trivial topological conducting orders, would be of deep and fundamental interest. Moreover this could also possibly lead to the discovery of further features which differ from the non interacting topological case \cite{KCM}.
Due to the fact that string orders have been measured \cite{Endal,hilker}, ultracold quantum systems \cite{bloch} represent an ideal platform to study the possible appearance of topological effects in presence of interaction. Moreover the impressive level of control achieved with such experimental setups has also allowed to trap ultracold particles with long range dipolar interaction \cite{Lahaye2009}. By means of such a platform several spin models \cite{yan,depaz,lepoutre} with spin-spin exchange processes induced by the dipolar interaction have been reproduced. At the same time when spin exchange is also associated to particles motion one gets an hybrid spin chain, namely the $t-J$ model \cite{zhang,dagotto,chao}.
\begin{figure}[]
\includegraphics[scale=0.9]{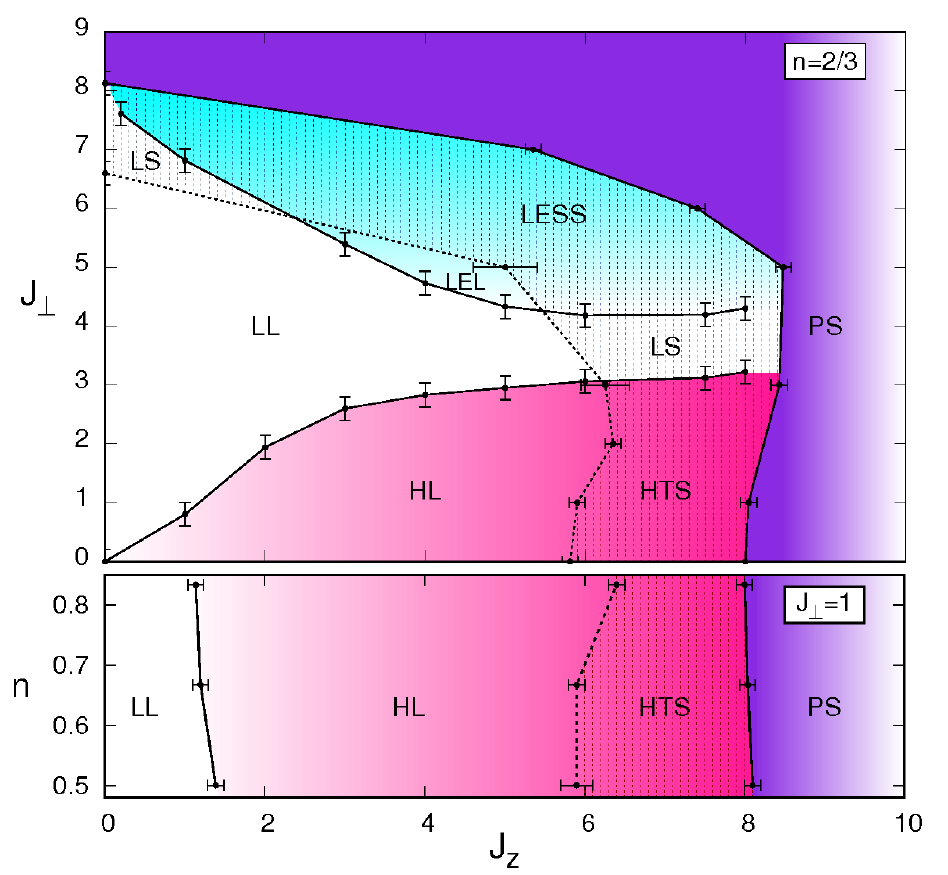}
\caption{DMRG phase diagram. \textit{Upper panel} Phase diagram at fixed density $n=2/3$ and $t=1$ as a function of $J_z$ and $J_\perp$. It consists of four phases: Luttinger liquid (LL, white area), Luther Emery liquid (LEL, cyan area), Haldane liquid (HL, pink area) and phase separation (PS, purple area). The solid lines correspond to $\Delta_s\sim 2\times 10^{-3}$ \cite{nota_DeltaS} for the HL-LL and LL-LEL transitions, and $K_c^{-1}\to0$ for the transition to PS. In each phase, the thick dashed line ($K_c=1$) identifies the crossover to the superconducting regime  $K_c>1$ (dashed area): Haldane liquid with dominant triplet superconductivity (HTS), Luther-Emery liquid with dominant singlet superconductivity (LESS) and Luttinger liquid with dominant either triplet or singlet superconductivity (LS). \textit{Lower panel} Phase diagram as a function of the density and $J_z$, at fixed $J_{\perp}=1$ and $t=1$.}
\label{Fig:PD}
\end{figure} 
This Hamiltonian has its own special relevance because it gives a proper description of quantum magnetism \cite{auerbach,lacroix} and high energy processes \cite{grusdt}. Furthermore since the interplay between hole motion and antiferromagnetism, peculiar of cuprate superconductors \cite{And} is properly captured by the $t-J$ Hamiltonian, the latter represents a fundamental model where high $T_c$ superconductivity can be studied \cite{ogata}. Importantly it has to be underlined that, since the $t-J$ model arises from the strong coupling limit of the Hubbard model, only a small portion of the phase diagram can be reliably investigated, namely the one where $J$ is isotropic and $J<<t$. However, thanks to the possibility to trap systems of ultracold fermionic polar molecules \cite{ni,aikawa,will,demarco} an anisotropic version of the $t-J$ model with independently tunable coupling constants can be achieved \cite{gorshkov}, thus allowing to explore the full phase diagram.

Motivated by such a possibility in this paper we  explore the intriguing interplay between superconductivity and topological orders occurring in the $t-J_z-J_\perp$ model. Our analysis based on bosonization technique \cite{Giam} and density-matrix-renormalization-group (DMRG) algorithm \cite{white} allows to derive a rich phase diagram as function of the antiferromagnetic anisotropy and the particle density. As shown in Fig. \ref{Fig:PD}, besides a phase separated (PS)
state, it amounts to a gapless Luttinger liquid (LL) phase and two spin gapped phases,  one with trivial and one with non trivial topological features. The latter is characterized by both a finite value of a string order parameter and by the appearance of degenerate fractionalized edge modes detected by the edge magnetization. Relevantly, by varying the anisotropy parameter $J_z$, we also find that superconducting orders can become dominant. Indeed in the spin gapped phase with non trival topology these manifest as leading triplet superconducting correlations, thus providing a first framework to realize topological superconductivity solely induced by interaction.

\paragraph{Model.}
As derived in \cite{gorshkov} polar molecules in the electronic and vibrational ground state with isolated rotational modes are captured by the following Hamiltonian

\begin{eqnarray}
{H}=&-&t\sum_{i,\sigma} \left (c^\dagger_{i,\sigma} c_{i+1,\sigma}+h.c.\right )+
\nonumber\\
&+&\sum_{i<j}\frac{1}{|i-j|^3} \Big[\frac{J_\perp}{2} (S^+_i S^-_j +S^-_iS^+_j)+J_z S_i^zS_j^z+
\nonumber\\
&+&Vn_in_j+Wn_iS^z_j\Big]\quad
\label{ham}
\end{eqnarray}
describing a system of $N=N_\uparrow+N_\downarrow$ (with $N_\uparrow=N_\downarrow$) fermionic particles loaded in $L$ sites, with total density $n=N/L$. In particular $c^\dagger_{i,\sigma}$ creates a fermion with dressed rotor state or, analogously, with spin state $\sigma$ in the $i$-site and $S^+_i=c^\dagger_{i,\uparrow}c_{i,\downarrow}$ , $S^z_i=(n_{i,\uparrow}-n_{i,\downarrow})/2$ are customarily defined as spin $1/2$ operators in a fermionic representation. Besides $t=1$ which fixes our energy scale and characterizes the hopping processes of a fermion tunneling in a nearest neighbor (NN) site, the other coupling constants $J_\perp, J_z, V$ and $W$ describe antiferromagnetic exchange in the $x-y$ plane and in the $z$ plane, density-density, and a density-spin interaction respectively. Furthermore, due to fact that eq. (\ref{ham}) can be realized with highly reactive molecules, double occupancies $(\uparrow\downarrow)$, are strictly forbidden. This aspect is taken into account by projecting the model eq. (\ref{ham}) onto the subspace with a vanishing number of doubly occupied sites, $H\rightarrow P H P$, with $P\doteq \prod_i (1-n_{i\uparrow} n_{i\downarrow})$, thus giving rise to a truncated local Hilbert space $(0, \uparrow, \downarrow)$.\\ For NN couplings eq. (\ref{ham}) has been intensively studied in different regimes. In particular for $J_\perp=J_z$, $V=-1/4$ and $W=0$ one recovers the well known $t-J$ model \cite{zhang,dagotto,chao}. Relevantly, the possibility to tune all the parameters has made reliable also the study of other cases. In particular, for $J_z=V=W=0$ enhanced superconductivity \cite{MMGH} and $d$-wave superfluidity \cite{kuns} have been found, whereas for $J_\perp=V=W=0$ superconducting behaviors \cite{leung,batista} and mesonic resonances \cite{GKBal} are expected. Nevertheless in the aforementioned regimes topological phases have not been predicted.\\ Here we study the more general situation where $V=W=0$ and both $J_\perp$ and $J_z$ are finite and can take different values thus describing an anisotropic $t-J$ model. Since in 1D couplings decaying like $|i-j|^{-\alpha}$ with $\alpha>1$ are not expected to generate new phases \cite{Giam} we consider the interactions limited to NN sites. In fact the inclusion of longer range couplings turns out to just modify the shape of the quantum phases but not their nature (see supplemental material \cite{SM}).

\paragraph{Bosonization.}	
In the above hypothesis, the model eq. (\ref{ham}) can be regarded as a Hubbard Hamiltonian with anisotropic Heisenberg interaction in the limit of infinite on-site repulsion $U$. This model has been studied within bosonization at finite $U$ both at \cite{JaMH} and away \cite{Dzal} from half-filling. In the second case the fundamental ground state features may be extracted by taking the limit $U\to\infty$ of the bosonization analysis in the hypothesis that stronger interaction is not capable to open further phases. One finds that depending on the value of the anisotropy $\delta\doteq J_z/J_{\perp}$ the system can be either in a gapless Luttinger liquid phase ($\delta<1$) or in a spin gapped phase ($\delta>1$). In the latter case, the specific value of the bosonic field reveals \cite{MoRo,barbiero1,DoMo} that the opening of the spin gap is associated uniquely to a non vanishing string parameter (see also below), thus displaying the appearance of a Haldane liquid (HL) phase. At the same time numerical studies \cite{MMGH}  have shown that the $\delta=0$ case supports the presence of a spin gapped Luther Emery liquid (LEL) phase not predicted by the above bosonization analysis. Thus, here we follow also an alternative route based on treating the kinetic term projected with $P$ as correlated hopping processes \cite{SM}. In this way we are able to predict the appearance of both the HL and the LEL phases for $\delta\neq 0$.
\\In each of the above phases, the actual value of the charge Luttinger parameter $K_c$, which we will properly define later, can be used to identify the regime where superconducting correlations become dominant \cite{Giam}. In particular the value $K_c=1$ characterizes the crossover to the superconducting regimes. As shown in Fig. \ref{Fig:PD} we find that $K_c>1$ can occur in all the possible conducting phases. More precisely we get that in the gapless LL phase both triplet (TS) and singlet (SS) superconducting orders can become dominant, describing a Luttinger superconductor (LS) regime. On the other hand, the gapped phases support the presence of only one type of superconductivity: SS in the LEL phase (LESS regime), and TS in the HL phase, thus describing an Haldane liquid in which a regime with with dominant triplet superconductivity appears (HTS regime). 

\paragraph{Topological features.} 
A bosonization analysis can show that the two spin gapped phases are associated to specific nonlocal order parameters defined as $O_{S/P}=\lim_{r\to\infty}O_{S/P}(r)$, with
\begin{equation}
O_S(r)=4 \langle S_j^z \prod_{l=j}^{j+r-1}e^{\imath2\pi S_l^z}S_{j+r}^z\rangle
\label{string}
\end{equation}
\begin{equation}
O_P(r)=\langle \prod_{l=j}^{j+r-1}e^{\imath 2\pi S_l^z}\rangle\hspace{2pt},
\label{parity}
\end{equation}
and called string and parity respectively. The string order parameter $O_S$ is nonzero in the whole HL phase, while it vanishes in the LL and LEL phases; whereas the parity $O_P$ is nonzero in the entire LEL phase and zero in the LL and HL phases \cite{SM}, \cite{barbiero1}. We point out that, at variance with the parity order, a hidden string order detected by a non vanishing $O_S$\cite{Nijs1989} is a typical signature of the topological nature of the corresponding Haldane phase \cite{Wen,PTBO,MDIR}. Thus, we expect that such phase hosts entangled fractionalized spins localized at the edges of an open chain, which average value differs from the bulk one: $< S_1^z>_\pm=-<S_L^z>_\pm \neq 0,\pm 1/2$. Here $\langle...\rangle_{\pm}$ denotes the expectation value taken on the two degenerate ground states $|\psi_{GS}\rangle_\pm $.
\\For $J_\perp=0$ the above topological features can be evaluated explicitly \cite{tJz}. The ground state has been discussed in \cite{batista}, upon recognizing that the particles must have alternated spins and thus can be replaced by spinless fermions. For $J_z>8 t$, phase separation occurs, where particles and empty sites are immiscible. Whereas for $J_z<8 t$ the ground state is conducting, and superconducting correlations are dominant for $J_z^{c}<J_z<8 t$\cite{Hal,QFYal}. Since the particles have alternated spin orientation, we observe that the phases must also be spin gapped. This is consistent with our previous bosonization analysis, where a spin gapped topological phase was identified for $\delta<1$. The result is also confirmed by the value of the string order parameter in such phase: $O_S(r)\xrightarrow[r\rightarrow \infty]{} n^2$ \cite{tJz}. Similarly one can calculate the fractional spin located at the edges, obtaining 
\begin{equation}
<S_1^z>_\pm =\pm \frac{n}{2}=-<S_L^z>_\pm \quad . \label{edge} 
\end{equation}
The subsequent numerical analysis will show that both topological properties hold qualitatively also in the non integrable case $J_{\perp}\neq0$, in a large portion of the phase diagram.

\paragraph{DMRG analysis.}
In order to study also the $J_{\perp}\neq0$ case and to validate the bosonization predictions, a priori reliable for weak interaction, we provide quasi exact DMRG results \cite{nota_DMRG}. The numerical phase diagram is shown in Fig. \ref{Fig:PD}, at fixed filling $n=2/3$ (upper panel) and fixed $J_{\perp}=1$ (lower panel). 
\begin{figure}[H]
\includegraphics[scale=0.25]{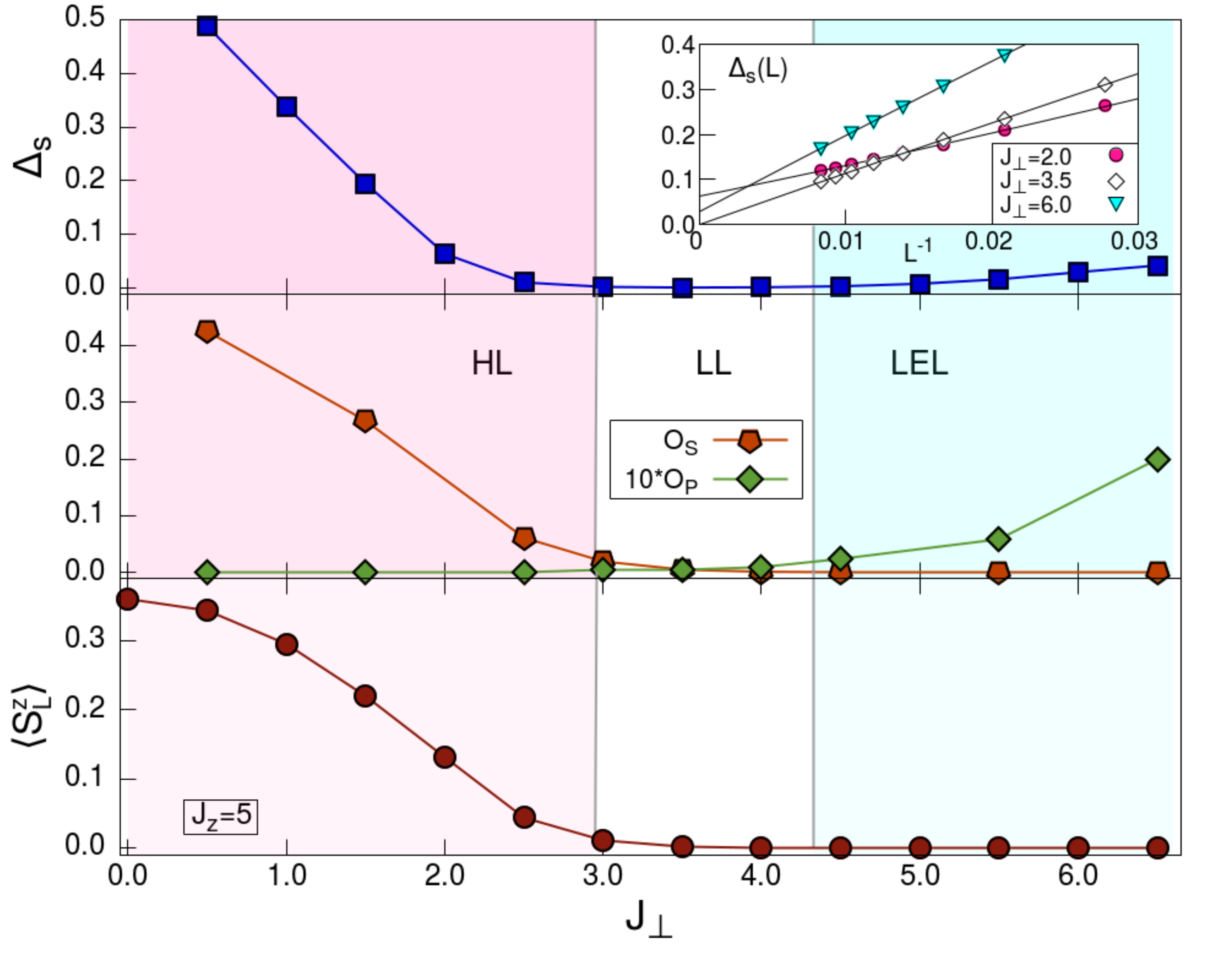}
\caption{\textit{Upper panel} Spin gap in the TDL at fixed $J_z=5$ and extrapolated by keeping $L$ up to 120. The inset shows examples of the finite size scaling in LL (white squares), LEL (cyan triangles) and HL (pink circles). $\Delta_s(L)$ has been obtained by using open boundary conditions, keeping up to 500 states and 5 finite size sweeps. \textit{Central panel} Nonlocal string (\ref{string}) and parity (\ref{parity}) order parameters obtained from finite size scaling of $O_S(L/2)$ and $[O_P(L/2-1)+O_P(L/2)+O_P(L/2+1)]/3$ computed for systems up to $L=48$. We used periodic boundary conditions (PBC) and keep up to 1200 states and 6 finite size sweeps. \textit{Lower panel} Edge magnetization $<S^z_L>$ on the last site for an unbalanced system with $N_{\uparrow}=N_{\downarrow}+1$. All the results are obtained by fixing $t=1$ and $n=2/3$.}
\label{Gap}
\end{figure}
The first fundamental quantity to properly capture all the miscible phases is the spin gap $\Delta_s=\lim_{L\rightarrow \infty} \Delta_s(L)$ , where $\Delta_s(L)=\lim_{L\rightarrow \infty}\left[E(N=L,S^z_{tot}=1)-E(N=L,S^z_{tot}=0)\right]$ and $E(N,S^z_{tot})$ is the ground state energy of a system with $N$ particles and total magnetization $S^z_{tot}=\sum_{i=1}^LS_i^z$. As shown in Fig. \ref{Gap} we find that for small $J_{\perp}$ a region with open spin gap is present. Once $J_{\perp}$ is increased, the competition between the two antiferromagnetic couplings generates a fully gapless LL phase. At the same time Fig. \ref{Gap} also makes evident that a further increase of $J_{\perp}$ allows for the appearance of another phase with $\Delta_s\neq 0$. This validates the bosonization predictions regarding the presence of two distinct regions with open spin gap. As shown in the central panel of Fig. \ref{Gap} the latter are each characterized by the non vanishing of one of the two nonlocal order parameters eqs. (\ref{string}) and (\ref{parity}). More precisely we obtain that for small $J_{\perp}$ hidden $z$-antiferromagnetism is favorable. This gives rise to a topological HL phase signaled by $O_S\neq 0$ (pink region in Fig. \ref{Fig:PD}). On the other hand for large $J_{\perp}$ the spin gap turns out to be associated to $O_P\neq 0$, thus identifying the trivial LEL phase (cyan region in Fig. \ref{Fig:PD}). As mentioned, the appearance of fractional edge modes is captured by the value of the edge magnetization. In the lower panel of Fig. \ref{Gap} we show that, even for $J_{\perp}\neq0$, $<S^z_L>\neq0$ remains finite only in the topological phase with non vanishing $O_S$. It approaches the value $n/2$ of eq. (\ref{edge}) in the integrable limit $J_\perp=0$ while reaching the asymptotic value $1/L$ \cite{note} in all the other phases.

\begin{figure}
\includegraphics[scale=0.55]{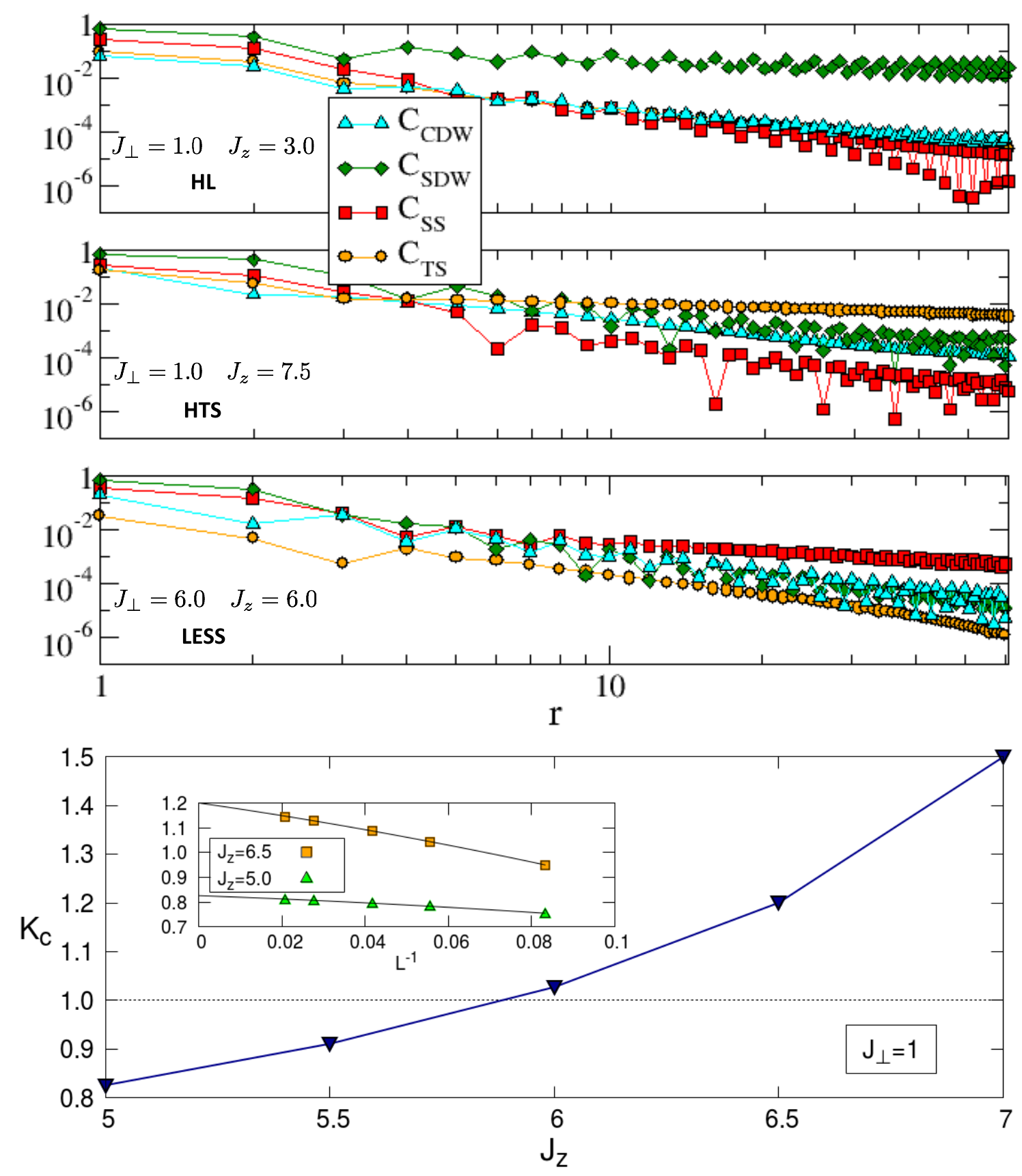}
\caption{\textit{Upper panel} Decay of the correlation functions in HL with $K_c<1$, HTS and LESS regimes. The correlations have been computed with OBC for a chain of length $L=120$, between the site $L/4$ and the site at distance $r$. \textit{Lower panel} Charge Luttinger parameter in the TDL at fixed $J_{\perp}=1$. The inset shows the finite size scaling in the two regions with $K_c<1$ and $K_c>1$. All the results are obtained by fixing $t=1$ and $n=2/3$}
\label{Fig:Kc}
\end{figure}

Moreover, for stronger values of the couplings the system undergoes a further phase transition entering in a region of phase separation. This is captured by $K_c^{-1}\to 0$ which signals a diverging value of the compressibility. As customary, we have extrapolated $K_c$ from the charge structure factor $S(q)=\frac{1}{L}\sum_{i,j}e^{\imath q(i-j)}\left(\langle n_i n_j\rangle-\langle n_i\rangle\langle n_j\rangle \right)$:
\begin{equation}
K_c=\lim_{q\to 0}\frac{\pi}{q}S(q) \hspace{2pt},
\end{equation}
in order to locate the transition line.\\ As already discussed, in each phase the value of $K_c$ identifies also the crossover to the regime in which superconducting correlations become dominant (dashed regions in Fig. \ref{Fig:PD}). By combining the procedure just explained with a finite size extrapolation (see lower panel of Fig. \ref{Fig:Kc}) we locate the corresponding transition line ($K_c=1$) reported in Fig. \ref{Fig:PD}. Moreover in order to enforce the results, in the upper panel of Fig. \ref{Fig:Kc} we have checked the power law decays of the relevant conducting orders in the different regions of the phase diagram. We evaluated the following correlation functions:
\begin{equation}
\begin{split}
&C_{SDW}(r)=\langle S_i^zS_{i+r}^z\rangle\\
&C_{CDW}(r)=\langle n_i n_{i+r}\rangle -\langle n_i\rangle\langle n_{i+r}\rangle\\
&C_{TS}(r)=\langle O_{TS}^{\dagger}(i)O_{TS}(i+r)\rangle\\
&C_{SS}(r)=\langle O_{SS}^{\dagger}(i)O_{SS}(i+r)\rangle
\end{split}
\end{equation}
with $O_{TS}^{\dagger}(i)=\frac{1}{\sqrt{2}}\left(c_{i,\uparrow}^{\dagger}c_{i+1,\downarrow}^{\dagger}+c_{i,\downarrow}^{\dagger}c_{i+1,\uparrow}^{\dagger}\right)$ and $O_{SS}^{\dagger}(i)=\frac{1}{\sqrt{2}}\left(c_{i,\uparrow}^{\dagger}c_{i+1,\downarrow}^{\dagger}-c_{i,\downarrow}^{\dagger}c_{i+1,\uparrow}^{\dagger}\right)$. In the upper panel of Fig. \ref{Fig:Kc}, we find that for small $J_z>J_{\perp}$, non superconducting correlations  ($C_{SDW}$) are the leading order in the topological HL phase. Whereas for larger $J_z$ values it is clearly seen that singlet and triplet superconductivity become the dominant orders in the trivial (LESS regime) and topological (HTS regime) phases respectively, in agreement with the behavior expected for $K_c>1$. Thus we have unambiguously demonstrated that in the model eq. (\ref{ham}) superconductivity can coexist with topological properties like fractionalized edge modes.

\paragraph{Conclusions.}
We derived the phase diagram of a generalized $t-J$ model in presence of spin anisotropy. Here the competition between hole motion and antiferromagnetic coupling gives rise to a rich phase diagram. The latter reveals the presence of a gapless Luttinger  liquid phase surrounded by large regions where the spin gap becomes finite. Moreover the study of correlation functions allows to notice how, among different conducting orders, superconductivity can become dominant. By means of nonlocal order parameters, we found that the spin gap is generated by two different mechanisms: either by virtual excitations of the vacuum composed by bounded fermions with antiparallel spins, thus captured by a parity operator; or by hidden antiferromagnetic order among particles with antiparallel spin , thus described by a string correlator. The latter scenario is associated to the presence of degenerate fractionalized edge states. Relevantly such topological order occurs also where superconducting correlations are dominant. Hence our results provide a fundamental microscopic description of topological superconductivity induced by interaction. They also open the way towards the observation of new properties of such topological matter, which are expected\cite{KCM} to drastically differ from those appearing in non interacting systems. In conclusion, it is worth underlying that all our results can be tested and reproduced by means of the ongoing experimental techniques involving polar molecules \cite{yan}. Indeed this platform only requires  in-situ probes to measure nonlocal order parameters \cite{Endal,hilker}, local magnetization \cite{parson} and density-density correlation to extrapolate the Luttinger constant \cite{schauss}.

\begin{acknowledgments}
{\it Acknowledgments:} 
The authors thank G. Japaridze and L. Santos for interesting discussions. L. B. acknowledges ERC Starting Grant TopoCold for financial support.
\end{acknowledgments}


\begin{thebibliography}{}
\bibitem{nota_DeltaS} Below that threshold the spin gap is considered equal to zero. This numerical error has been estimated upon extrapolating to the thermodynamic limit and by comparison with the typical values of $\Delta_s$ in the gapped phases, that are one or two order of magnitudes larger (see Fig. \ref{Gap}).
\bibitem{nota_DMRG} In our numerical simulations we have used both open boundary conditions (OBC) and periodic boundary conditions (PBC), as specified in the captions of the figures. In both cases we have chosen the number of DMRG states and sweeps in such a way as to ensure convergence within a truncation error of the order or smaller than $10^{-6}$. In particular, in case of PBC we have checked that translational invariance is accurately reproduced.
\bibitem{note} This value is due to the fact that in order to see edge magnetization the 2-fold degeneracy of the ground state has to be broken. We do this by adding a further particle which preserves the string order and does not affect the value of the spin gap.
\end{thebibliography}
\end{document}